\documentstyle[11pt,nature]{article}
\begin{document}
\heading{Detection of Large Scale Structure in a $\bf B < 17^{m}$
Galaxy Redshift Survey.}

\author{A. Ratcliffe*, T. Shanks*, A. Broadbent*, Q.A. Parker\dag,}
{F.G. Watson\dag, A.P. Oates\ddag, R. Fong* \& C.A. Collins\S}
{*Physics Dept., Univ. of Durham, South Road, Durham, DH1 3LE, U.K.}
{\dag Anglo-Australian Observatory, Coonabarabran, NSW 2357,
Australia.}
{\ddag Royal Greenwich Observatory, Madingley Road, Cambridge, CB3 0HA,
U.K.}
{\S Astrophysics Group, Liverpool-John-Moores Univ., Liverpool, L3 3AF,
U.K.}

\noindent{\bf We report on results from the Durham/UKST Galaxy Redshift
Survey where we have found large scale ``cellular'' features in the
galaxy distribution. These have spatial 2-point correlation function
power significantly in excess of the predictions of the standard cold
dark matter cosmological model$^{1}$, supporting the previous
observational results from the APM survey$^{\bf 2,3}$. At smaller
scales, the 1-D pairwise galaxy velocity dispersion is measured to be
$\bf 387^{+96}_{-62}$ kms$^{-1}$ which is also inconsistent with the
prediction of the standard cold dark matter model$^{1}$. Finally, the
survey has produced the most significant detection yet of large scale
redshift space distortions due to dynamical infall of galaxies$^{4}$.
An estimate of $\bf \Omega^{0.6}/b = 0.55 \pm 0.12$ is obtained which
is consistent either with a low density Universe or a critical density
Universe where galaxies are biased tracers of the mass.}

The principal aims of the Durham/UKST Galaxy Redshift Survey are to
investigate the structure and dynamics of the Universe on scales
1-100$h^{-1}$Mpc (where $h$ is Hubble's constant in units of 100
kms$^{-1}$ Mpc$^{-1}$). The survey was constructed by measuring the
redshifts for 1 in 3 of the galaxies from the Edinburgh/Durham Southern
Galaxy Catalogue$^{5}$ (EDSGC) to $b_{J} \simeq 17^{m}$. The resulting
survey contains $\sim 2500$ redshifts, covers a $\sim 20^{\circ} \times
75^{\circ}$ contiguous area of the sky at the South Galactic Pole and
probes to a depth of $> 300h^{-1}$Mpc with a median depth of $\sim
150h^{-1}$Mpc. The total volume of space surveyed is $\sim 4 \times
10^{6}h^{-3}$Mpc$^{3}$, approximately six times that of the updated
Southern Sky Redshift Survey$^{6}$, due to our $\sim 2.5^{m}$ fainter
limit.

Figure 1 shows the distribution of the galaxies in right ascension and
velocity for each of four, $5^{\circ}$, declination slices. The survey
gives the striking impression that the galaxy distribution is
``cellular'' or ``bubble-like'' on 50-100$h^{-1}$Mpc scales. The most
noticeable structure is the low density region lying between 0 and
90$h^{-1}$Mpc surrounded by long ``walls'' of galaxies. This structure
is present in the slices at $-30^{\circ}$, $-35^{\circ}$ and
$-40^{\circ}$ and was referred to as the Sculptor Void in earlier,
shallower surveys$^{7,8}$. In the most northerly slice at $-25^{\circ}$
there is evidence that we are seeing the top of this structure and that
it is indeed a ``cell''. Figure 2 shows the histogram of galaxy number
with distance which has several large peaks, also indicating the
presence of large scale structure. However, it should be noted that at
least one of these peaks (eg. at $\sim 90h^{-1}$Mpc) does not follow
the 128$h^{-1}$Mpc periodic pattern previously claimed$^{9}$ along the
North--South Galactic Pole axis which intersects our survey at $\sim
0^{h} 54^{m}$, $-27.5^{\circ}$, crossing the nearer and further
Sculptor superclusters. Figure 1 clearly reveals a galaxy distribution
that is more complex than any simple periodic pattern.

To test the significance of these large scale features, we have made a
statistical investigation of the galaxy distribution in the Durham/UKST
survey by calculating the redshift space 2-point correlation
function$^{10}$, $\xi(s)$. This is a measure of the excess probability
of finding two galaxies in given volume elements at separations $s$
over what is expected for a homogeneous galaxy distribution. In figure
3a we compare the Durham/UKST $\xi(s)$ with that from other optical
redshift surveys$^{3,11,12}$. On small scales ($< 10h^{-1}$Mpc) we see
that there is good agreement between all the surveys. On large scales
($> 10h^{-1}$Mpc) we see that our new estimate is consistent with the
detection of large scale power from the APM-Stromlo survey but less
consistent with the previous Durham surveys. This inconsistency is
thought to be partly statistical and partly due to the previous use of
a simple unweighted estimator which, although having minimum errors,
was systematically biased against the detection of large scale power
(Ratcliffe \et, in preparation). Our new estimate is also consistent
with that found from the QDOT infrared redshift survey$^{13}$. Figure
3b shows the Durham/UKST $\xi(s)$ with the results from mock catalogues
drawn from two sets of cold dark matter$^{1}$ (CDM) N-body
simulations$^{14,15,16}$ ; standard CDM with $\Omega h = 0.5$, $b=1.6$
(SCDM) and CDM with $\Omega h = 0.2$, $b=1$ and a cosmological constant
($\Lambda \ne 0$) to ensure a spatially flat cosmology (LCDM), where
$\Omega$ is the mean mass density of the Universe and $b$ is the linear
bias parameter relating the galaxy and underlying matter distributions,
$(\Delta\rho/\rho)_{g} = b (\Delta\rho/\rho)_{m}$. The mock catalogues
were constructed with the same angular and radial selection functions
as the Durham/UKST survey. On small scales ($< 10h^{-1}$Mpc) both CDM
models agree well with the data. On large scales ($> 10h^{-1}$Mpc) SCDM
shows no significant large scale power whereas LCDM shows significant
power out to $\sim 30h^{-1}$Mpc. The data therefore shows significant
excess power at $\sim$ 15-30$h^{-1}$Mpc over SCDM ($> 3 \sigma$). LCDM
is more consistent with the data at these scales, although even this
model produces too little power at the $2 \sigma$ level. This rejection
of SCDM in our new survey is consistent with the findings from the APM
correlation functions$^{2,3}$.

Inferring galaxy distances from redshifts results in the distortion of
the clustering pattern because of the non-Hubble component of galaxy
velocities. This anisotropy along the line of sight can be used to
estimate some important cosmological parameters describing the dynamics
of the Universe$^{4,10}$. Figure 4a shows a contour plot of the 2-point
correlation function, $\xi_{v}(\sigma,\pi)$, where $\sigma$ and $\pi$
are the perpendicular and parallel separations to the line of sight,
respectively. On small, non-linear scales the contours are elongated
parallel to the line of sight caused by the rms virial velocities of
galaxies. Using standard techniques$^{10}$ to model this smearing of
$\xi$ by the peculiar velocities, we find $<v^{2}>^{\frac{1}{2}} =
387^{+96}_{-62}$ kms$^{-1}$ at perpendicular separations $\sigma <
2h^{-1}$Mpc. Assuming that simple biasing models apply$^{17}$, this is
inconsistent with the SCDM prediction$^{1}$ of $\sim 1000$ kms$^{-1}$
but nearer the LCDM prediction$^{1}$ of $\sim 600$ kms$^{-1}$. On
large, linear scales the contours are compressed parallel to the line
of sight caused by the infall of galaxies into overdense regions. We
use two independent methods to estimate $\Omega^{0.6}/b$ as shown in
figure 4b. Firstly, using the spherical harmonic moments$^{18}$
($\widetilde{\xi}$) of $\xi_{v}(\sigma,\pi)$ to measure the shape of
the contours of constant $\xi$, we find $\Omega^{0.6}/b = 0.55 \pm
0.12$ in the region 9-23$h^{-1}$Mpc. Secondly, using the enhancement in
redshift space$^{4}$ of the volume integral of $\xi$ ($J_{3}$) over the
real space one, we find $\Omega^{0.6}/b = 0.48 \pm 0.18$, quoting the
result at $10h^{-1}$Mpc, which is representative. Our measured values
of $\Omega^{0.6}/b$ can be compared with other optical values of
$\Omega^{0.6}/b$ estimated from redshift space distortions, namely
$0.48 \pm 0.12$ (Loveday \et, manuscript submitted), $0.77 \pm
0.16^{19}$ and $0.5 \pm 0.25^{20}$. All of these values agree within
2$\sigma$ of our results. Thus taking two fiducial values for $b$,
$b=1$ implies $\Omega = 0.37 \pm 0.15$ and $b=2$ implies $\Omega = 1.17
\pm 0.31$. This result tends to argue against an unbiased critical
density, $\Omega = 1$, Universe, but favours a biased, $\Omega = 1$,
Universe or an unbiased, low $\Omega$, Universe.

Thus our $\Omega^{0.6}/b$ result agrees with the SCDM and LCDM models
but, since they both predict $\Omega^{0.6}/b \sim$ 0.4-0.6, we cannot
discriminate between them. However, we have shown that SCDM
underpredicts our 2-point correlation function results at large scales
and overpredicts the 1-D pairwise velocity dispersion at small scales.
Therefore, overall, our results argue for a model with a density
perturbation spectrum more skewed towards large scales, such as LCDM.

\acknowledgements

We are grateful to the staff at the UKST and AAO for their assistance
in the gathering of the observations. S.M. Cole, C.M. Baugh and V.R.
Eke are thanked for useful discussions and supplying the CDM
simlations. PPARC are thanked for allocating the observing time via
PATT and for the use of the STARLINK computer facilities.
   
\references

\natref{1}{Davis, M., Efstathiou, G., Frenk, C.S. \& White,
S.D.M.}{\aj}{{\bf 292}, 371--394}{(1985)}

\natref{2}{Maddox, S.J., Efstathiou, G., Sutherland, W.J. \& Loveday,
J.}{\mn}{{\bf 242}, 43p--47p}{(1990)}

\natref{3}{Loveday, J., Efstathiou, G., Peterson, B.A. \& Maddox,
S.J.}{\aj {\em $\;\;$Lett.}}{{\bf 400}, L43--L46}{(1992)}

\natref{4}{Kaiser, N.}{\mn}{{\bf 227}, 1--21}{(1987)}

\natref{5}{Collins, C.A., Heydon-Dumbleton, N.H., \& MacGillivray,
H.T.}{\mn}{{\bf 236}, 7p--12p}{(1988)}

\natref{6}{Marzke, R.O., Geller, M.J., da Costa, L.N. \& Huchra,
J.P.}{{\em Astron. J.}}{{\bf 110}, 477-501}{(1995)}

\natref{7}{da Costa, L.N., Pellegrini, P.S., Davis, M., Meiksin, A.,
Sargent, W.L. \& Tonry, J.L.}{\aj {\em $\;\;$Suppl. Ser.}}{{\bf 75},
935--964}{(1991)}

\natref{8}{Fairall, A.P. \& Jones, A.}{{\em Publs. Dept. Astr. Cape
Town}}{{\bf 10}}{(1988)}

\natref{9}{Broadhurst, T.J., Ellis, R.S., Koo, D.C. \& Szalay,
A.S.}{\nat}{{\bf 343}, 726--728}{(1990)}

\natref{10}{Peebles, P.J.E.}{{\em `The Large-Scale Structure of the
Universe'},}{Princeton University Press, Princeton, US}{(1980)}

\natref{11}{Shanks, T., Bean, A.J., Efstathiou, G., Ellis, R.S., Fong,
R., \& Peterson, B.A.}{\aj}{{\bf 274}, 529--533}{(1983)}

\natref{12}{Shanks, T., Hale-Sutton, D., Fong, R. \& Metcalfe, N.}{\mn}
{{\bf 237}, 589--610}{(1989)}

\natref{13}{Saunders, W., Frenk, C.S., Rowan-Robinson, M., Efstathiou,
G., Lawrence, A., Kaiser, N., Ellis, R.S., Crawford, J., Xia, X.-Y. \&
Parry, I.}{\nat}{{\bf 349}, 32--38}{(1991)}

\natref{14}{Efstathiou, G., Davis, M., Frenk, C.F. \& White, S.D.M.}
{\aj {\em $\;\;$Suppl. Ser.}}{{\bf 57}, 241--260}{(1985)}

\natref{15}{Couchman, H.M.P.}{\aj {\em $\;\;$Lett.}}{{\bf 368},
L23--L26}{(1991)}

\natref{16}{Gazta\~{n}aga \& Baugh, C.M.}{\mn}{{\bf 273},
1p--6p}{(1995)}

\natref{17}{Bardeen, J.M., Bond, J.R., Kaiser, N. \& Szalay,
A.S.}{\aj}{{\bf 304}, 15--61}{(1986)}

\natref{18}{Hamilton, A.J.S.}{\aj {\em $\;\;$Lett.}}{{\bf 385},
L5--L8}{(1992)}

\natref{19}{Peacock, J.P. \& Dodds, S.J.}{\mn}{{\bf 267},
1020--1034}{(1994)}

\natref{20}{Lin, H.}{{\em Ph.D. Thesis},}{University of
Harvard}{(1995)}

\natref{21}{Parker, Q.A. \& Watson, F.G.}{in {\em `Wide Field
Spectroscopy and the Distant Universe'}, $35^{th}$ Herstmonceux
Conference, Cambridge, UK, (eds. Maddox, S.J. \& Arag\'{o}n-Salamanca,
A.),}{World Scientific Publishing, 33--39}{(1995)}

\natref{22}{Peterson, B.A., Ellis, R.S., Efstathiou, G.P., Shanks, T.,
Bean, A.J., Fong, R., \& Zen-Long, Z.}{\mn}{{\bf 221},
233--255}{(1986)}

\natref{23}{Metcalfe, N., Fong, R., Shanks, T. \& Kilkenny, D.}{\mn}
{{\bf 236}, 207--234}{(1989)}

\natref{24}{Hamilton, A.J.S.}{\aj}{{\bf 417}, 19--35}{(1993)}

\natref{25}{Efstathiou, G.}{in {\em `Comets to Cosmology'}, $3^{rd}$
IRAS Conference, London, UK, (ed. Lawrence, A.),}{Springer-Verlag,
312--319}{(1988)}

\natref{26}{Saunders, W., Rowan-Robinson, M. \& Lawrence, A.} {\mn}{{\bf
258}, 134--146}{(1992)}

\newpage

\begin{flushleft}
{\bf Figure 1.}
\end{flushleft}
Redshift cone-plots of the galaxy distribution in the four contiguous
$5^{\circ}$ declination strips from the Durham/UKST redshift survey,
(a), (b), (c) and (d), respectively. Large ``cellular'' structures in
the galaxy distribution on scales of 50-100$h^{-1}$Mpc are clearly
seen. The redshifts of these galaxies were obtained over the three year
period 1991-1994 using the 1.2m UK Schmidt Telescope at Siding Spring,
Australia, in conjunction with the fibre-coupled spectroscopy system
FLAIR$^{21}$. The completeness is $> 75\%$ to the nominal magnitude
limit of $b_{J} = 17.0^{m}$. Our magnitude limit varies slightly
(0.1$^{m}$-0.2$^{m}$) from field to field but this is exactly taken
into account in all our statistical analysis. A comparison of $150$
galaxies with published velocities$^{7,8,22,23}$ indicates that our
measurements have a negligible offset and are accurate to $\pm 150$
kms$^{-1}$ (Ratcliffe \et, in preparation).

\begin{flushleft}
{\bf Figure 2.}
\end{flushleft}
The histogram of galaxy number with distance, $n(r)$, in the
Durham/UKST survey. The dashed curve shows how a random and homogeneous
distribution would appear given our angular/radial selection functions,
sampling rate and galaxy luminosity function from the Durham/UKST
survey (Ratcliffe \et, in preparation). The arrows indicate where the
128$h^{-1}$Mpc ``spikes'' in the galaxy distribution$^{9}$ should
appear in this region of the sky. In this larger angle survey the
galaxy distribution clearly appears more complex than any simple
periodic pattern. There are two strong peaks at $\sim 90$ and $\sim
170h^{-1}$Mpc signifying ``walls'' in the galaxy distribution. There is
possible evidence for a third such feature at $\sim 270h^{-1}$Mpc.

\newpage

\begin{flushleft}
{\bf Figure 3.}
\end{flushleft}
In (a) we compare the Durham/UKST 2-point correlation function,
$\xi(s)$, with that from other optical redshift surveys$^{3,11,12}$.
$\xi(s)$, was calculated using the weighted estimator$^{24,25}$ which
gave no systematic offset and the minimum variance in the CDM mock
catalogues. On this plot the error bars for the Durham/UKST survey were
calculated using the 1$\sigma$ variance from the LCDM mock catalogues.
In (b) we compare with a SCDM model and a LCDM model. The shaded areas
are the 1$\sigma$ confidence regions on an individual mock catalogue
for the above two models of CDM using the aforementioned optimal
estimator and weighting combination so that a direct comparison between
the data and models can be made. We have also estimated the integral
constraint to see if it would cause any systematic offset in $\xi(s)$
on large scales and find this to be a negligible effect.

\begin{flushleft}
{\bf Figure 4.}
\end{flushleft}
In (a) we show the contours of constant 2-point correlation function
where the effects of small and large scale redshift space distortions
are clearly seen. Solid lines denote $\xi > 1$ with $\Delta \xi = 1$,
short dashed lines denote $0 < \xi < 1$ with $\Delta \xi = 0.1$ and
long dashed lines denote $\xi < 0$ with $\Delta \xi = 0.1$. For
reference, the lines with $\xi=1$ and $\xi=0$ are in thick bold and the
regularly spaced thin bold lines show an isotropic correlation function
for comparison. A representative error bar is also plotted. In this
plot $\xi_{v}(\sigma,\pi)$ was calculated using a single pair weighting
of galaxies$^{10}$ because this weighting produced less noise in
$\xi_{v}(\sigma,\pi)$. Our consistent use of this different weighting
scheme has no effect on the either method of estimating
$\Omega^{0.6}/b$ (Ratcliffe, \et in preparation). In (b) we show our
estimated values of $\Omega^{0.6}/b$ for two methods$^{18,4}$ as a
function of separation $r$. The solid line shows the maximum likelihood
fit for $\Omega^{0.6}/b$ to the first method of estimation
($\widetilde{\xi}$) in the region 9-23$h^{-1}$Mpc and the dashed lines
denote the $1 \sigma$ confidence intervals of this fit. The error bars
on these independent points come from the LCDM mock catalogues. The
error bar of the estimated point at $\sim 10h^{-1}$Mpc from the second
method ($J_{3}$) is representative of the errors from this method and
was calculated from the variance between the 4 quadrants of the survey.
Only one error bar is shown because of the non-independence of the
points in this method. Both of these sets of error bars include the
contribution from cosmic variance. We have estimated $\xi$ in real
space using two methods involving the inversion of the projected
correlation function$^{26}$ (Ratcliffe \et, in preparation) and find
consistent results.

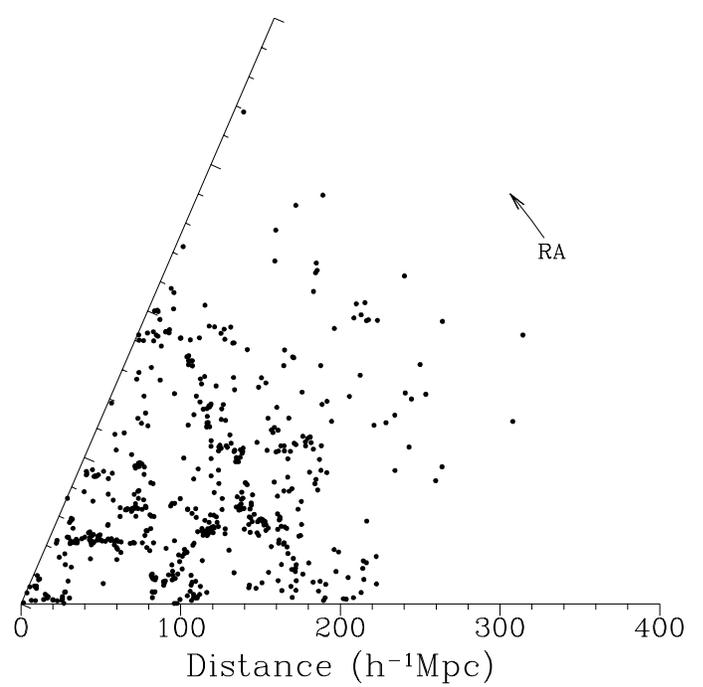
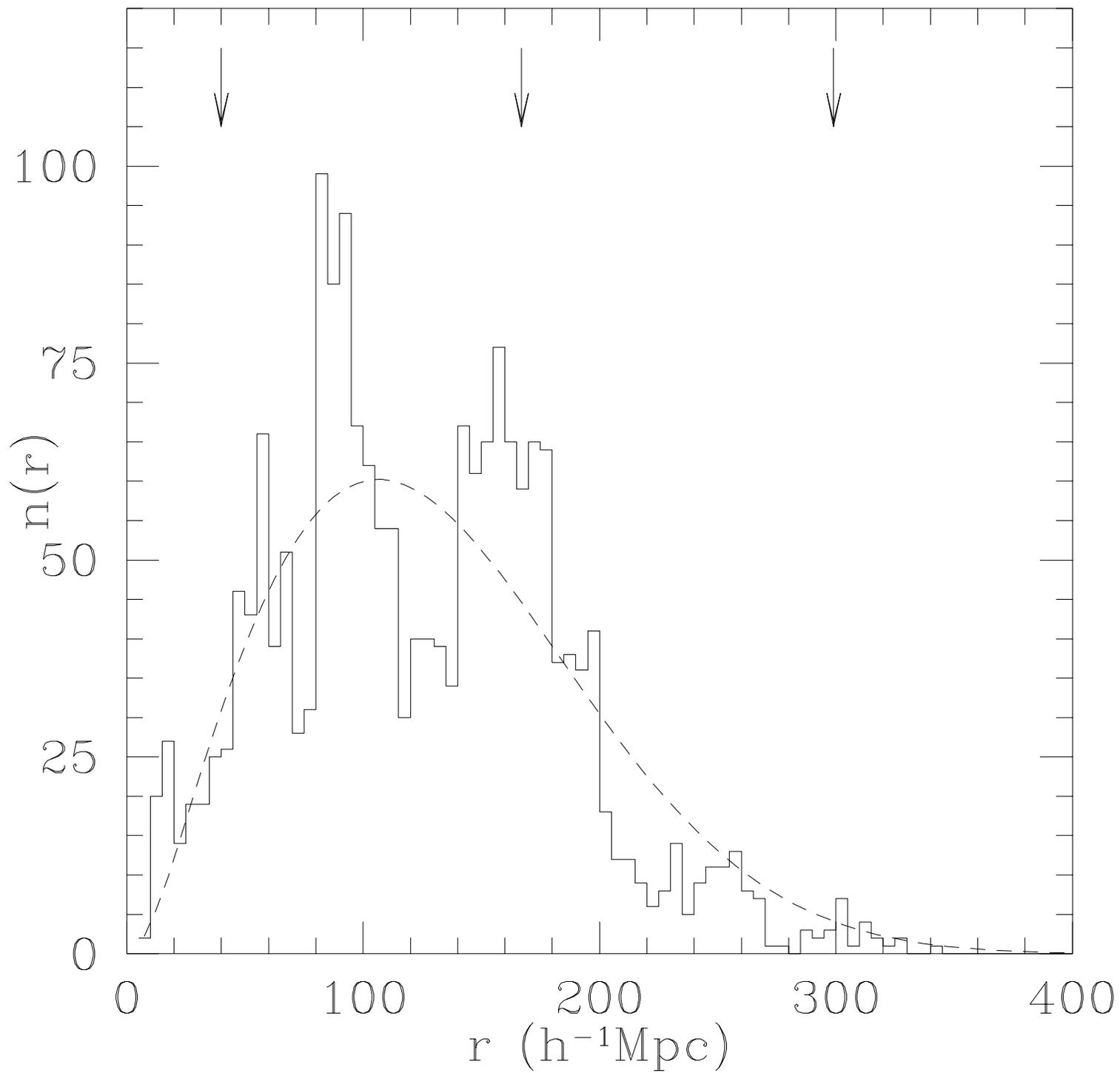
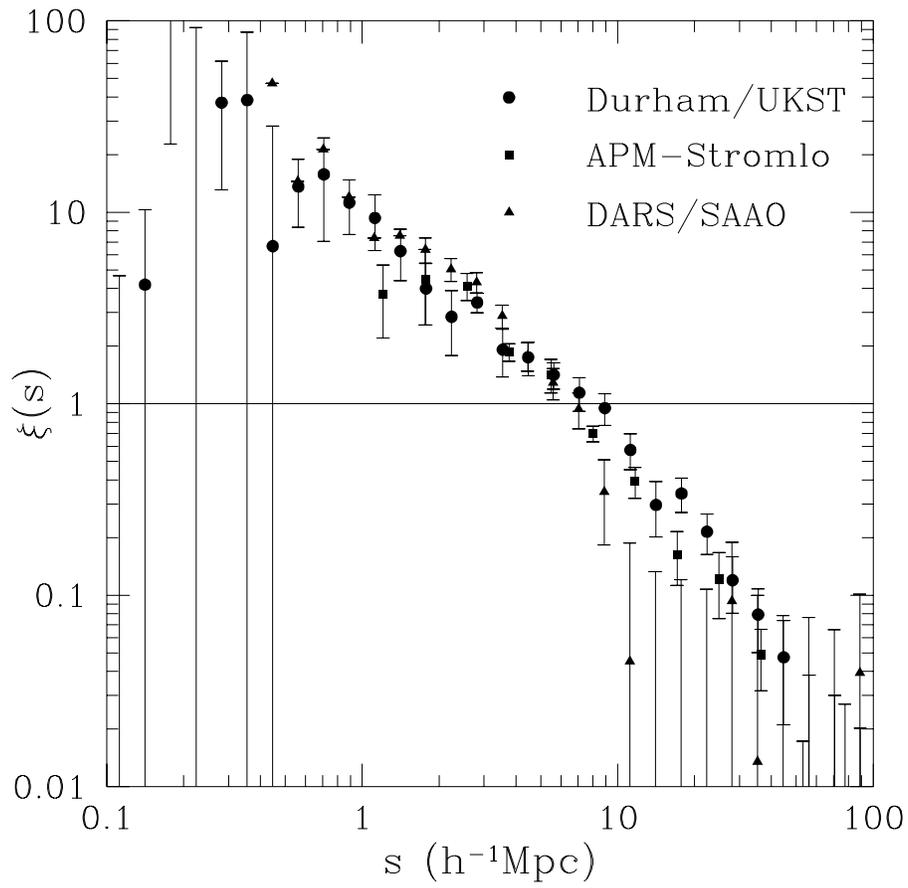
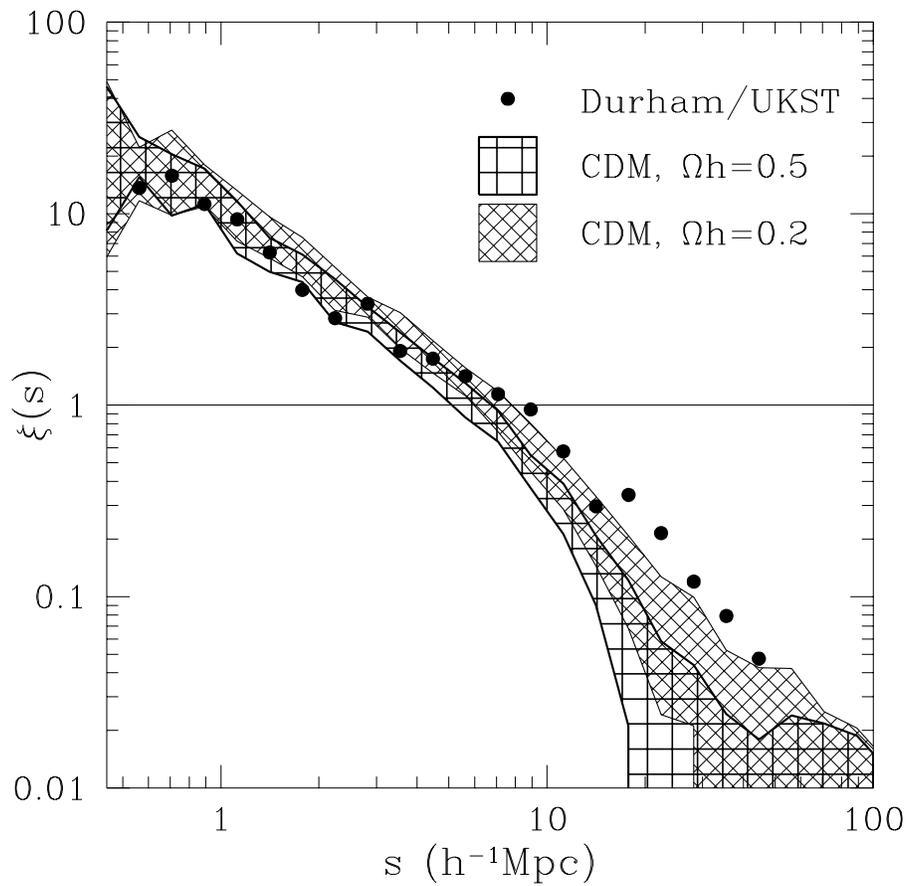
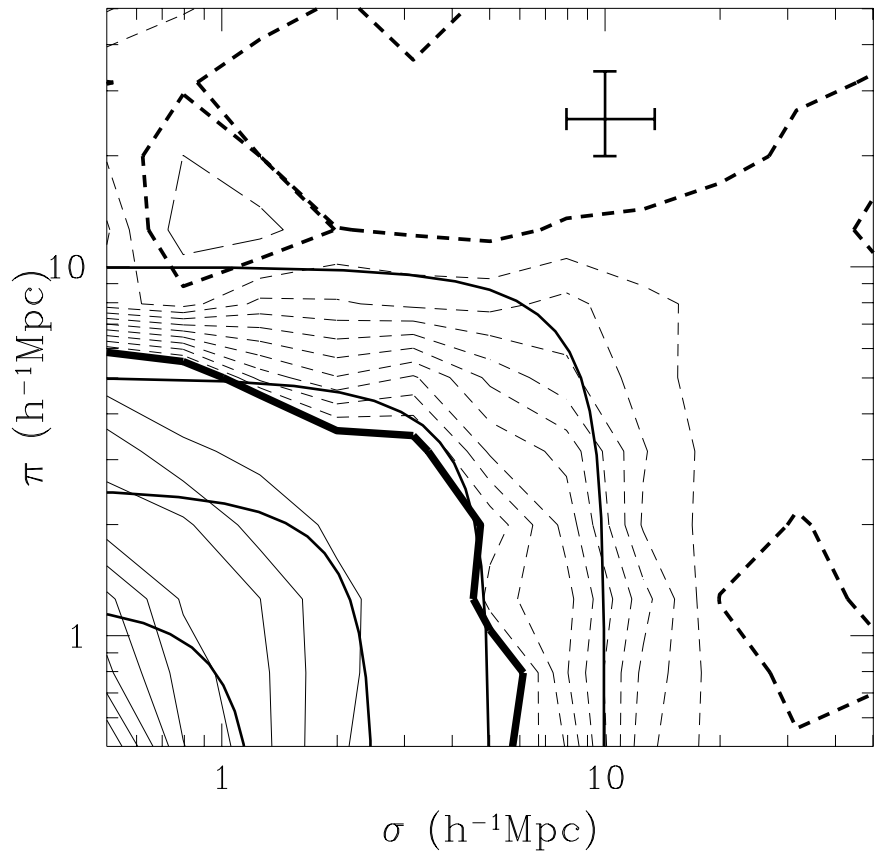
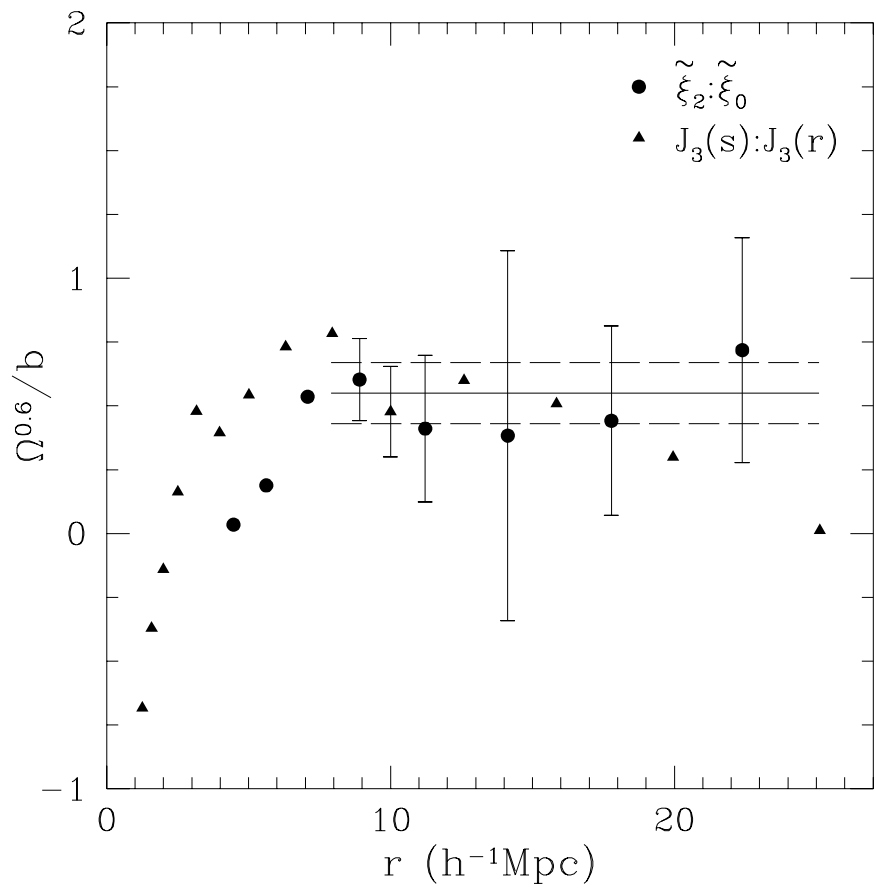
\end{document}